\documentclass[reprint,showpacs,preprintnumbers,amsmath,amssymb,aps,prb,]{revtex4-1}

\usepackage{graphicx}
\usepackage{dcolumn}
\usepackage{bm}
\usepackage{color}
\usepackage{enumerate}
\usepackage{placeins}

\definecolor{darkorange}{RGB}{255,140,0}

\def\Kd{\mathrm{d}}

\def\KAS{\mathbb{S}}
\def\KPermG{\mathcal{P}}
\def\Ke#1{e^{#1}}
\def\KS{\mathcal{S}}
\def\KG{\mathcal{G}}

\def\KT{\Gamma}
\def\Kcbr#1{\left[ #1\right ]}
\def\Kbr#1{\left( #1\right )}
\def\Kb#1{\overline{#1}}
\def\Kfd{\partial}
\def\Ktd{\mathrm{d}}
\def\Kdmu#1{\mathrm{d}\mu_{#1}}
\def\P#1{_{#1}}
\def\Kph{\cdots}
\def\Kcirc{\circ \Pi_{G} \circ}
\def\Kwedge{\ast  \Upsilon_G \ast   }

\begin{document}

\title{Schwinger-Dyson Renormalization Group}

\author{Kambis Veschgini$^1$}
\email{k.veschgini@thphys.uni-heidelberg.de}
\author{Manfred Salmhofer$^{1,2}$}
\email{m.salmhofer@thphys.uni-heidelberg.de}
\affiliation{
$^1$Institut f\" ur Theoretische Physik, Universit\" at Heidelberg, D-69120 Heidelberg, Germany\\
$^2$Mathematics Department, University of British Columbia, Vancouver, B.C., Canada V6T1Z2
}

\date{\today}

\begin{abstract}
\noindent
We use the Schwinger-Dyson equations as a starting point to derive renormalization flow equations. We show that Katanin's scheme arises 
as a simple truncation of these equations. We then give the full renormalization group equations up to third order in the irreducible vertex. Furthermore, we show that to the fifth order, there exists a functional of the self-energy and the irreducible four-point vertex whose saddle point is the solution of Schwinger-Dyson equations.
\end{abstract}

\pacs{05.10.Cc,05.30.Fk,71.10.-w} %
                              
\maketitle

\section{Introduction}

\noindent
Renormalization group (RG) and Schwinger-Dyson equation (SDE) hierarchies 
are widely used methods to study the correlation functions of
quantum field theories. The SDE are a set of integral equations\cite{Dyson,Schwinger}
obtained, e.g., by integration by parts in the functional integral 
representation. The RG approach (for reviews, see e.g.\ Refs.\
\onlinecite{Berges02,SalmhoferHonerkamp_ProgTheor,MSHMS})
 starts by introducing a scale parameter
and a modification of the theory that regularizes, i.e. smoothes
out singularities, in the propagator. This regularizing effect
(often corresponding to a localization in position space)
is the main reason for the good mathematical properties of 
RG equations. The RG equation is a functional differential equation
which becomes a hierarchy of equations in the usual expansion in the 
fields. Instead of a self-consistency, it describes a flow. 
It is a natural idea to use such flow equations also to 
solve self-consistency equations, as an alternative to iterative solutions. 
Moreover, it is useful to make the relation between the two
as explicit as possible. In the first part of this paper, 
we formulate RG equations based on the SDE hierarchy, 
and then 
we relate Katanin's truncation \cite{PhysRevB.70.115109} 
of the RG hierarchy for the one-particle irreducible vertex functions
to a particular truncation of the RG derived from the SDE. 
We then also exhibit the higher-order terms in the SDE-RG.

The full SDE hierarchy encodes all analytic and 
combinatorial properties of the vertex functions, 
and retains the symmetries of the original action. 
Truncations of this hierarchy usually violate
Ward identities and conservation laws. 
In the RG approach, the same problem arises, 
but in addition, the regularization
may violate some symmetries explicitly, 
so that the restoration of Ward identities in the limit
where the regulator is removed requires proof. 
The most important example of this are theories with 
local gauge symmetries, e.g.\ QED\cite{KK1,KK2}.
A general theory of conserving approximations was developed by Baym and Kadanoff \cite{PhysRev.124.287,PhysRev.127.1391} in the context of many-body theory, 
and later also used 
in high-energy quantum field theory.\cite{CJT} 
The essential feature there is the Luttinger-Ward (LW)  functional, 
which expresses the grand canonical potential 
as a function of the bare vertex and the full propagator 
of the theory. The field equations are obtained by a 
stationarity condition as the propagator is varied. 
Similar constructions using the self-energy as the variational parameter instead of the propagator were introduced by Potthoff.\cite{Potthoff:2005,Potthoff:2012}
It is a natural question whether there is  a 
scale-dependent variant of this functional, 
in which both the full scale-dependent propagator  
and the effective two-particle vertex (instead of the bare one) appear. 
By iteration of the RG equations in their integral form, 
such an equation can be 
generated as an expansion in powers of the effective two-particle vertex
(in a procedure generalizing the derivations in Ref.\ \onlinecite{SHML}). 
If generated this way, the expansion, however, involves 
integrals over intermediate scales, similarly to the 
Brydges-Kennedy formula\cite{BK}, which provides an explicit solution to 
Polchinski's equation. 
Another generalization of the LW functional was given by the Lund group where the bare interaction is replaced by the screened interaction using the Bethe-Salpeter equation. The resulting functional is variational in both parameters.\cite{PhDHindgren,PhysRevB.69.195102}
In Section \ref{sec:functional} we derive from the SDE equations 
a functional of the self-energy and the irreducible four-point vertex, 
which is local in the RG flow parameter, at least up to the fifth order 
in the effective vertex. The stationary points of this functional satisfy the SDE. 
\section{Schwinger-Dyson equations and renormalization group}
\label{sec:katanin}
Consider a lattice fermion system described by Grassmann fields $\psi$,
$\Kb{\psi}$ and the action
\begin{equation}
\KS[\psi,\Kb{\psi}] = -(\Kb{\psi},C^{-1} \psi) - V\left[\psi,\Kb{\psi}\right]\;,
\end{equation}
where $C$ is the propagator of the non-interacting system and $V$ is a
two body interaction of the general form
\begin{equation}
V\Kcbr{\psi,\Kb{\psi}} = \frac14
\sum_{\begin{matrix}
\scriptstyle x_1,x_2,\,
\scriptstyle x_1',x_2' 
\end{matrix}}
v\P{x_1',x_2',x_1,x_2}\Kb{\psi}\P{x_1'}\Kb{\psi}\P{x_2'}\psi\P{x_2}\psi\P{x_1} \;.
\end{equation}
In applications, the labels $x$ are often composed of several variables, in momentum space with a single
particle basis one has $x=(k_0,\mathbf{k},\sigma)$, where $k_0$ is the
Matsubara frequency, $\mathbf{k}$ the is the momentum and $\sigma$ denotes 
the spin orientation. Depending on the representation there might be prefactors like volume or temperature which are not shown here. $v$ is antisymmetric under independent 
exchange of its first two and last two arguments.
The bilinear form $(f,g)$ is defined as the sum $\sum_x f(x) g(x)$.

The generating functional of the connected Green functions is given by
\begin{equation}
\label{eq:GeneratorG}
\KG\Kcbr{\eta,\Kb{\eta}} = -\ln \int \Kdmu{C}  \Ke{V\Kcbr{\psi,\Kb{\psi}}}
\Ke{ \Kbr{\Kb{\eta},\psi}+\Kbr{\Kb{\psi},\eta} }\;,
\end{equation}
where $\Kdmu{C} := {\mathcal N} \prod_x \Kd \psi\P{x} \Kd \Kb{\psi}\P{x} e^{(\Kb{\psi},C^{-1} \psi) }$, with a normalization constant ${\mathcal N}$. 
The connected $m$-particle Green function is obtained from the generator $\KG$ by differentiating 
with respect to the sources $\eta,\Kb{\eta}$ and evaluating for vanishing sources:
\begin{equation}
\begin{split}
& G^{(2m)}\P{ x_1,\dots,x_m;x_1',\dots,x_m' } =\\
&\; (-1)^m \frac{\Kfd^{2m}\KG\Kcbr{\eta,\Kb{\eta}} }
{\Kfd \Kb{\eta}\P{x_1} \dots \Kfd \Kb{\eta}\P{x_m}\Kfd \eta\P{x_m'}\dots \Kfd \eta\P{x_1'}}
\Bigg|_{\eta=\Kb{\eta}=0}=\\ 
&\; -\langle   \psi\P{x_1}\dots\psi\P{x_m}\Kb{\psi}\P{x_m'}\dots\Kb{\psi}\P{x_1'} \rangle_c \; .
\end{split}
\end{equation}
Here $\langle \Kph  \rangle_c$ stands for the connected average of the expression between the brackets.
The {\em effective action} is the Legendre transform of
$G\Kcbr{\eta,\Kb{\eta}}$
\begin{equation}
\label{eq:1PIGenerator}
\KT\Kcbr{\psi,\Kb{\psi}} = \Kbr{\Kb{\eta},\psi} + \Kbr{\Kb{\psi},\eta} + G\Kcbr{\eta,\Kb{\eta}}\;,
\end{equation}
with
$
\psi = - \frac{\Kfd \KG}{\Kfd \Kb{\eta}}$ %
and
$\Kb{\psi} =  \frac{\Kfd \KG}{\Kfd \eta}\;.
$
It generates the one-particle irreducible (1PI) Green functions
$\Gamma^{(2m)}\P{ x_1,\dots,x_m;x_1',\dots,x_m' }$.

\begin{figure*}[]
\includegraphics{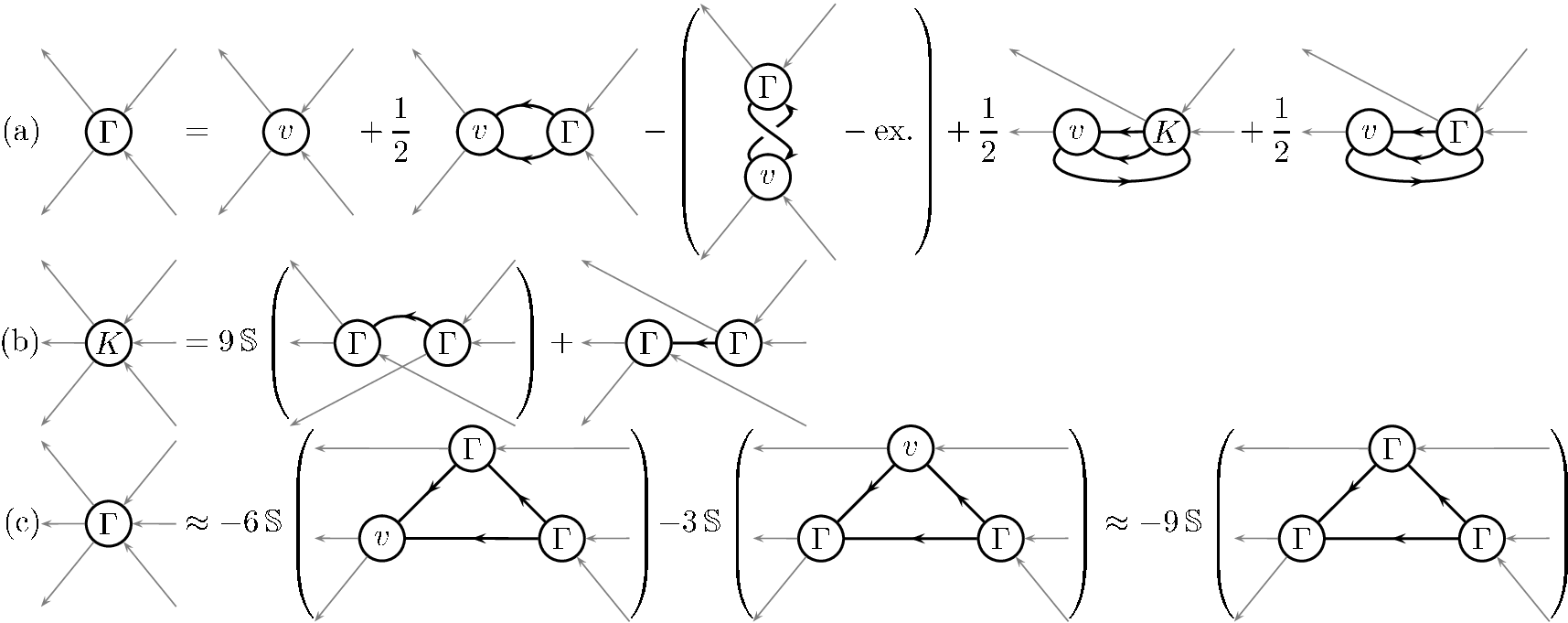}
\caption{ The diagrammatic representations of the SD eq.~\eqref{eq:Gamma4Exact} is given in (a), (b) represents eq.~\eqref{eq:K} and (c) corresponds to eq.~\eqref{eq:DSG6} }.

\label{fig:SDG4}
\end{figure*}

By integration by parts
\begin{equation}
\label{eq:IntegrationByParts}
\int \Kdmu{C} \psi_x F[\psi,\Kb{\psi}] = - \sum_y C\P{x,y} \int \Kdmu{C} \frac{\Kfd}{\Kfd \Kb{\psi}_y} F[\psi,\Kb{\psi}] \;,
\end{equation}
every correlation function obeys a Schwinger-Dyson equation
\begin{equation}
\label{eq:SD}
\begin{split}
&\int \Kdmu{C} \psi\P{x_1}\dots \psi\P{x_m} \Kb{\psi}\P{x_m'}\dots \Kb{\psi}\P{x_1'} \Ke{V\Kcbr{\psi,\Kb{\psi}}}  =  \\
&-\;\sum_y C\P{x_1,y} \int \Kdmu{C} \frac{\Kfd }{\Kfd \Kb{\psi}\P{y} } \psi\P{x_2}\dots \psi\P{x_m} \Kb{\psi}\P{x_m'}\dots \Kb{\psi}\P{x_1'} \Ke{V\Kcbr{\psi,\Kb{\psi}}}  \;.
\end{split} 
\end{equation}
The correlation function on the right hand side is in general disconnected,
but can be expressed by standard cumulant formulas
in terms of the connected Green functions,
which are in turn given by a standard expansion in trees
that have the full propagator $G:=G^{(2)}$ associated to lines
and the 1PI vertex functions to the vertices
.
For $m=1$, eq.\  (\ref{eq:SD}) gives an equation for $G$.
\begin{equation}
\label{eq:DSG}
\begin{split}
&G\P{x,x'} = C\P{x,x'}
-\sum_{z_1\cdots z_4} \, C\P{x,z_1}
G\P{z_4,z_2} v\P{z_1,z_2,z_3,z_4} G\P{z_3,x'}\\
&-\frac{1}{2}
\sum_{\begin{matrix}
\scriptstyle z_1\cdots z_4\\ 
\scriptstyle y_1\cdots y_4
\end{matrix}} \Big(
 C\P{x,z_1}
v\P{z_1,z_2,z_3,z_4} G\P{z_3,y_1} G\P{z_4,y_2}
\Gamma^{(4)}\P{y_1,y_2,y_3,y_4}\cdot\\
&\qquad\qquad\qquad G\P{y_4,z_2} G\P{y_3,x'}   \Big)
\;,
\end{split}
\end{equation}
After a rewriting
in terms of the self-energy $\Sigma = C^{-1}-G^{-1}$
one obtains
$\Sigma = C^{-1}-G^{-1}$ is given by
\begin{equation}
\label{eq:Self-Energy}
\begin{split}
&\Sigma\P{x,x'} =
-\sum_{z_2, z_4} \,
G\P{z_4,z_2} v\P{x,z_2,x',z_4} -\\
&\frac{1}{2} 
\sum_{\begin{matrix}
\scriptstyle z_2\cdots z_4\\ 
\scriptstyle y_1\cdots y_3
\end{matrix}}
v\P{x,z_2,z_3,z_4} G\P{z_3,y_1} G\P{z_4,y_2}
\Gamma^{(4)}\P{y_1,y_2,x',y_4} G\P{y_4,z_2}  
\;.
\end{split}
\end{equation}
The SD equation for $m=2$ gives the 4-point vertex 
(2-particle vertex) as
\begin{equation}
\label{eq:Gamma4Exact}
\begin{split}
&\Gamma^{(4)}\P{x_1,x_2;x_1',x_2'} = 
 v\P{x_1,x_2;x_1',x_2'}\\
&+
\frac12 \sum_{z_1\cdots z_4} v\P{x_1,x_2;z_1,z_2}  G\P{z_1,z_3} G\P{z_2,z_4} \Gamma^{(4)}\P{z_3,z_4;x_1',x_2'} \\
&- \Kbr{
\sum_{z_1\cdots z_4} v\P{x_1,z_1;x_3,z_2}  G\P{z_4,z_1} G\P{z_2,z_3} \Gamma^{(4)}\P{z_3,x_2;z_4,x_2'}
-\Kbr{x_3\leftrightarrow x_4}}\\
&+
\frac12 \sum_{z_1\cdots z_6} v\P{z_1,x_1;z_2,z_3} G\P{z_3,z_5} G\P{z_2,z_4} G\P{z_6,z_1}  K\P{z_4,z_5,x_2,z_6,x_1',x_2'}\\
&+
\frac12 \sum_{z_1\cdots z_6} v\P{z_1,x_1;z_2,z_3} G\P{z_3,z_5} G\P{z_2,z_4} G\P{z_6,z_1}  \Gamma^{(6)}\P{z_4,z_5,x_2,z_6,x_1',x_2'}
\;,
\end{split}
\end{equation}

where 
\begin{equation}
\label{eq:K}
\begin{split}
& K\P{x_1,x_2,x_3;x_1',x_2',x_3'} :=\\ 
& \quad 9\, \KAS_3 
\sum_{z_1,z_2} \Gamma^{(4)}\P{x_2,x_3;x_1',z_1}  G\P{z_1,z_2} \Gamma^{(4)}\P{x_1,z_2;x_2',x_3'}\\
&+\sum_{z_1,z_2} \Gamma^{(4)}\P{x_1,x_2;x_1',z_1}  G\P{z_1,z_2} \Gamma^{(4)}\P{z_2,x_3;x_2',x_3'}
\;,
\end{split}
\end{equation}
and the antisymmetrization operator  $\KAS_m$ 
projects functions $f\P{x_1,\cdots;x_1',\dots,x_m'}$ 
to their totally antisymmetric part
\begin{equation}
\begin{split}
&\KAS_m \, 
f\P{x_1,\dots,x_m;x_1',\cdots,x_m'} = \\
&\quad \frac{1}{(m!)^2}\sum_{\pi,\pi' \in \KPermG_m} 
\mathrm{sgn}(\pi)\mathrm{sgn}(\pi')
f\P{x_{\pi(1)},...,x_{\pi(m)},x'_{\pi'(1)},...,x'_{\pi'(m)} }\;.
\end{split}
\end{equation}
The second term in eq.~\eqref{eq:K} cancels a contribution from the first term which would otherwise lead to a reducible diagram in eq.~\eqref{eq:Gamma4Exact}.
In a graphical representation where a vertex $f$ is depicted by
\begin{equation}
\label{eq:vertex}
f\P{x_1,\dots,x_m;x_1',\dots,x_m'} \simeq
\vcenter{\hbox{\includegraphics{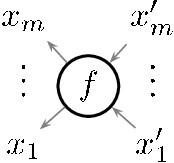}}}\;.
\end{equation}
and where heavy lines represent propagators $G$,
\begin{equation}
\label{eq:prop}
G\P{x_1;x_1'} \simeq
\vcenter{\hbox{\includegraphics{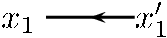}}}\;.
\end{equation}
eq.~\eqref{eq:Gamma4Exact} takes the form shown in Fig.~\ref{fig:SDG4}.

If $C=C_\Lambda$ depends on a parameter $\Lambda$, 
the SD equations \eqref{eq:Self-Energy} and \eqref{eq:Gamma4Exact} 
determine $\Lambda$-dependent self-energies $\Sigma_\Lambda$
and two-particle vertices $\Gamma^{(4)}_\Lambda$, etc.
We assume that $C$ depends differentiably on $\Lambda$ and that 
the bare interaction $V$ remains independent of $\Lambda$. 
Following the standard convention in fRG studies of condensed-matter systems, 
we arrange things such that for some value $\Lambda_0$ of $\Lambda$ 
(the ``starting scale''), 
the vertex functions are given by the bare ones, and the full correlation functions
are recovered as $\Lambda \to 0$. 
Moreover, we assume that $\Lambda$ is introduced in a way that 
has a regularizing effect, so that $\Sigma_\Lambda$
and $\Gamma^{(2m)}_\Lambda$ are differentiable functions of $\Lambda$ as well 
and derivatives with respect to $\Lambda$ can be exchanged with the summations
occurring in the SDE. Note that this is an assumption on the {\em solution} of the hierarchy, 
which will in general contain singular functions in the limit $\Lambda \to 0$, 
hence checking it is important and nontrivial. However, for the standard 
momentum space cutoff RG, it has been proved,\cite{salmCMP98,msbook}
and this proof extends to any RG flow that imposes a sufficient regularization on $C$, 
in particular the temperature RG flow\cite{HonerkampSalmhofer_Tflow}, 
flows with a frequency cutoff or the $\Omega$-regularization.\cite{HusemannSalmhofer2009}
Thus the assumption is satisfied in a large class of flows, 
for which the SDE hold at every scale $\Lambda$. 

Our above choice to make $C$, but not $V$, depend on $\Lambda$ is natural here 
because we want to draw a connection between SDE and standard RG flows. 
One can think of many other useful ways in which a parameter $\Lambda$ could 
be introduced in the SDE, also in the interaction (or only there). A natural way
to check the differentiability assumption is then to 
truncate the SDE hierarchy at successive levels, and within each truncation verify 
the differentiability  conditions by analysis of the right-hand side of the flow equation.

The derivative of eq.~\eqref{eq:Gamma4Exact} 
with respect to $\Lambda$ (denoted here by a dot)
then gives rise to terms on the right hand side where 
only propagators are differentiated, and ones where 
$\dot \Gamma^{(4)}_\Lambda$ appears. 
For instance, the derivative of the particle-particle term
\begin{equation}
\label{eq:Gamma4pp}
\frac12 \sum_{z_1\cdots z_4} v\P{x_1,x_2;z_1,z_2} 
 G\P{z_1,z_3} G\P{z_2,z_4} \Gamma^{(4)}\P{z_3,z_4;x_1',x_2'}
\end{equation}
is
\begin{equation}
\label{eq:Gamma4ppdot}
\begin{split}
&\frac12 \sum_{z_1\cdots z_4} v\P{x_1,x_2;z_1,z_2} 
 \frac{\Kd }{\Kd \Lambda} \Kbr{G\P{z_1,z_3} G\P{z_2,z_4}} \cdot \Gamma^{(4)}\P{z_3,z_4;x_1',x_2'}
\\
& +\frac12 \sum_{z_1\cdots z_4} v\P{x_1,x_2;z_1,z_2} 
  G\P{z_1,z_3} G\P{z_2,z_4} \cdot \dot{\Gamma}^{(4)}\P{z_3,z_4;x_1',x_2'}
\;.
\end{split}
\end{equation}
It is possible to eliminate $v$ and $\dot{\Gamma}^{(4)}$ from the right hand side of eq.~\eqref{eq:Gamma4ppdot} by substituting $v$ from eq.~\eqref{eq:Gamma4pp} and iterating
eq.~\eqref{eq:Gamma4ppdot}.  
This results in
\begin{equation}
\label{eq:Gamma4Katanin}
\begin{split}
&\dot{\Gamma}^{(4)}\P{x_1,x_2;x_1',x_2'}
= \\
& \frac12 \sum_{z_1\cdots z_4} \Gamma^{(4)} \P{x_1,x_2;z_1,z_2} 
 \Kbr{  \frac{\Kd }{\Kd \Lambda} G\P{z_1,z_3} G\P{z_2,z_4}}  \Gamma^{(4)}\P{z_3,z_4;x_1',x_2'} \\
 &-(\text{ph.}-\text{ex.})+ \mathcal{O}  \Kbr{\Gamma^{(4)}}^3
\;.
\end{split}
\end{equation}
A similar procedure applied to eq.~\eqref{eq:Self-Energy} gives
\begin{equation}
\label{eq:SelfEnergyKatanin}
\begin{split}
\dot{\Sigma} \P{x_1,x_1'} &= 
-\sum_{z_2, z_4} \,
\dot{G}\P{z_4,z_2} v\P{x,z_2,x',z_4} -\\
&\frac{1}{2} \frac{\Kd}{\Kd \Lambda}
\sum_{\begin{matrix}
\scriptstyle z_2\cdots z_4\\ 
\scriptstyle y_1\cdots y_3
\end{matrix}}
v\P{x,z_2,z_3,z_4} G\P{z_3,y_1} G\P{z_4,y_2}
\Gamma^{(4)}\P{y_1,y_2,x',y_4} G\P{y_4,z_2} \\
&=
- \sum_{z_2, z_4} \,
\dot{G}\P{z_4,z_2} v\P{x,z_2,x',z_4} 
+ \mathcal{O}  \Kbr{\Gamma^{(4)}}^2 \;.
\end{split}
\end{equation}
Eqs.\ \eqref{eq:SelfEnergyKatanin} and \eqref{eq:Gamma4Katanin} are the renormalization equations in
the Katanin scheme\cite{PhysRevB.70.115109}. 
In the first term in \eqref{eq:SelfEnergyKatanin}, $v$ could be replaced by $\Gamma^{(4)}$
up to orders $(\Gamma^{(4)})^2$, but as it stands, this term can be directly integrated, 
hence combined naturally with the resummation of the four-point function implied by 
keeping only one of the three terms in 
\eqref{eq:Gamma4Katanin}. 
This is at the basis of recovering selfconsistent ladder summations.\cite{PhysRevB.70.115109,SHML}
Because $\dot G = G \dot \Sigma G + S$, where $S$ is the single-scale propagator appearing 
in the standard RG equations for the irreducible vertex functions,\cite{SalmhoferHonerkamp_ProgTheor,MSHMS} we see that 
by substituting for $v$ in \eqref{eq:SelfEnergyKatanin}
by reinserting the SDE for $\Gamma^{(4)}$,
the standard 
1PI equation for the self-energy
\begin{equation}\label{standardself}
\dot{\Sigma} \P{x_1,x_1'} = 
-\sum_{z_2, z_4} \,
S\P{z_4,z_2} \Gamma^{(4)} \P{x,z_2,x',z_4} 
\end{equation}
is obtained when terms of third and higher order in $\Gamma^{(4)}$ are dropped. 
Within the 1PI RG hierarchy, \eqref{standardself} has no additional terms
of higher order in $\Gamma^{(4)}$.

\section{Higher order contributions to the self-consistent flow equations}
\label{sec:corrections}
Higher order contributions can be computed in a similar way by taking into account  
the higher order terms in the SD equations.

The six-point vertex in eq.~\eqref{eq:Gamma4Exact} can itself be expressed in terms of the interaction $v$, the four-point, the six-point and the eight-point vertices. At lowest order one obtains from eq.\eqref{eq:SD} for $m=3$,
\vbox{
\begin{equation}
\label{eq:DSG6}
\begin{split}
& \Gamma^{(6)}\P{x_1,x_2,x_3;x_1',x_2',x_3'} \approx 
\KAS \Big( \\
& -6 \sum_{z_1\cdots z_6} v\P{x_1,x_2;z_1,z_2}  \Gamma^{(4)}\P{z_3,x_3;z_4,x_3'}
\Gamma^{(4)}\P{z_5,z_6;x_1',x_2'}G\P{z_1,z_5} G\P{z_2,z_3} G\P{z_4,z_6} -\\
&3
\sum_{z_1\cdots z_6} \Gamma^{(4)}\P{x_1,x_2;z_1,z_2}  v\P{z_3,x_3;z_4,x_3'}
\Gamma^{(4)}\P{z_5,z_6;x_1',x_2'}G\P{z_1,z_5} G\P{z_2,z_3} G\P{z_4,z_6} \Big)\\
& \approx -9\KAS
 \sum_{z_1\cdots z_6}  \Gamma^{(4)}\P{x_1,x_2;z_1,z_2}  \Gamma^{(4)}\P{z_3,x_3;z_4,x_3'}
\Gamma^{(4)}\P{z_5,z_6;x_1',x_2'} \times \\
&\qquad\qquad\qquad\qquad\qquad\qquad G\P{z_1,z_5} G\P{z_2,z_3} G\P{z_4,z_6} 
\;.
\end{split}
\end{equation}}
In the last step we have replaced $v$ by $\Gamma^{(4)}+\mathcal{O}\Kbr{\Gamma^{(4)}}^2$ according to eq.~\eqref{eq:Gamma4Exact}.
Following this procedure we obtain an equation $\Gamma^{(4)}-v=\dots$ where the right hand side consists of diagrams involving both the vertex $\Gamma^{(4)}$ and $v$. In this case $v$ can be eliminated from the right hand side by the means of iteration. 
This leads to a self-consistent equation $\Gamma^{(4)}-v=\Theta$,
as follows. 
Denote the particle-particle bubble propagator by $\Pi$
\begin{equation}
	(\Pi_G)\P{x_1,x_2;x_1',x_2'} := G\P{x_1,x_1'}G\P{x_2,x_2'}\;,
\end{equation}
and the particle-hole bubble propagator by $\Upsilon$
\begin{equation}
	(\Upsilon_G)\P{x_1,x_2;x_1',x_2'} := G\P{x_1,x_2'}G\P{x_1',x_2}\;,
\end{equation}
and define
\begin{equation}
	(f \circ g)\P{x_1,x_2;x_3,x_4} := \sum_{z_1,z_2} f\P{x_1,x_2,z_1,z_2} g\P{z_1,z_2,x_3,x_4} 
\end{equation}
and
\begin{equation}
	(f \ast g)\P{x_1,x_2;x_3,x_4} := \sum_{z_1,z_2} f\P{z_1,x_2,z_2,x_4} g\P{x_1,z_1,x_3,z_2} \;.
\end{equation}
In this notation
\begin{equation}
\label{eq:theta4}
\begin{split}
\Gamma^{(4)}&-v = \Theta := \KAS \Big\{ \frac12 \Gamma^{(4)} \Kcirc \Gamma^{(4)} -2 \Gamma^{(4)} \Kwedge \Gamma^{(4)}  \\
& -\frac14 \Gamma^{(4)} \Kcirc \Gamma^{(4)}\Kcirc \Gamma^{(4)}\\
& -2 \Gamma^{(4)} \Kwedge \Gamma^{(4)}\Kwedge \Gamma^{(4)}\\
& +\frac18  \Gamma^{(4)} \Kcirc \Gamma^{(4)}\Kcirc \Gamma^{(4)} \Kcirc \Gamma^{(4)} \\
& -2 \Gamma^{(4)} \Kwedge \Gamma^{(4)}\Kwedge \Gamma^{(4)} \Kwedge \Gamma^{(4)} \\
& -4 Q(G,\Gamma^{(4)})
\Big\}+\mathcal{O} \Kbr{\Gamma^{(4)}}^5\;.
\end{split}
\end{equation}
At this order, up to the last term, $\Theta$ consist of particle-particle and 
particle-hole ladder diagrams. The last term $Q(G,\Gamma^{(4)})$ is given by
\begin{equation}
\begin{split}
(Q(G &,\Gamma^{(4)}))\P{x_1,x_2;x_3,x_4} := \\ &\sum_{z_1 \cdots z_{12}} \Gamma^{(4)}\P{x_1,z_1;z_2,z_3}\Gamma^{(4)}\P{z_4,x_2;z_5,z_6}  \Gamma^{(4)}\P{x_7,z_8;x_3,z_9}\Gamma^{(4)}\P{z_{10},z_{11};z_{12},x_4} \cdot \\
& G\P{z_5,z_1} G\P{z_3,z_{10}} G\P{z_2,z_7} G\P{z_9,z_4} G\P{z_6,z_{11}} G\P{z_{12},z_8}\;.
\end{split}
\end{equation}
Eq.~\eqref{eq:theta4} is
interesting by itself and will be used in the next section to construct a functional whose stationary points are solutions of the SD equations.

The flow equation for the vertex $\Gamma^{(4)}$ is now given by 
$\dot{\Gamma}^{(4)}=\frac{\Kd}{\Kd \Lambda} \Theta$. Derivatives of $\Gamma^{(4)}$ which appear on 
the right hand side can be eliminated by iterating the result. 
The interaction $v$  does not appear in the flow equation, since it was assumed to be independent of $\Lambda$. It serves as the initial condition for the integration of $\dot{\Gamma}^{(4)}$. In the limit $\Lambda \to \infty$ where all fluctuations are suppressed $\Gamma^{(4)}=-v$. 
Although eq.~\eqref{eq:theta4} has a rather simple structure, 
the process of resubstituting $\dot{\Gamma}^{(4)}$ when it appears
on the right hand side 
mixes and proliferates the terms. 
The fourth order corrections are too long to be presented here. 
Up to third order, the flow equation is given by

\begin{widetext}
\vbox{
\begin{equation}
\begin{split}
\dot{\Gamma}\P{x_1,x_2;x_3,x_4}= \KAS_{x_1,x_2;x_3,x_4} \big(& \sum_{z_1\cdots z_{4}}\Gamma\P{x_1,x_2;z_{1},z_{2}}\Gamma\P{z_{3},z_{4};x_3,x_4}\dot{G}\P{z_{1},z_{3}}G\P{z_{2},z_{4}}
-4\sum_{z_1\cdots z_{4}}\Gamma\P{z_{1},x_1;x_3,z_{2}}\Gamma\P{x_2,z_{3};z_{4},x_4}\dot{G}\P{z_{4},z_{1}}G\P{z_{2},z_{3}}\\
&+4\sum_{z_1\cdots z_{8}}\Gamma\P{z_{1},x_1;x_3,z_{2}}\Gamma\P{z_{3},x_2;z_{4},z_{5}}\Gamma\P{z_{6},z_{7};z_{8},x_4}G\P{z_{8},z_{1}}G\P{z_{2},z_{3}}G\P{z_{4},z_{6}}\dot{G}\P{z_{5},z_{7}}\\
&+8\sum_{z_1\cdots z_{8}}\Gamma\P{z_{1},x_1;x_3,z_{2}}\Gamma\P{z_{3},x_2;z_{4},z_{5}}\Gamma\P{z_{6},z_{7};z_{8},x_4}G\P{z_{4},z_{1}}G\P{z_{2},z_{6}}G\P{z_{8},z_{3}}\dot{G}\P{z_{5},z_{7}}\\
&+8\sum_{z_1\cdots z_{8}}\Gamma\P{z_{1},x_1;x_3,z_{2}}\Gamma\P{z_{3},x_2;z_{4},z_{5}}\Gamma\P{z_{6},z_{7};z_{8},x_4}G\P{z_{4},z_{1}}G\P{z_{2},z_{6}}\dot{G}\P{z_{8},z_{3}}G\P{z_{5},z_{7}}\\
&+2\sum_{z_1\cdots z_{8}}\Gamma\P{x_1,z_{1};z_{2},z_{3}}\Gamma\P{z_{4},x_2;z_{5},z_{6}}\Gamma\P{z_{7},z_{8};x_3,x_4}\dot{G}\P{z_{6},z_{1}}G\P{z_{2},z_{4}}G\P{z_{3},z_{8}}G\P{z_{5},z_{7}}\\
&-4\sum_{z_1\cdots z_{8}}\Gamma\P{x_1,z_{1};z_{2},x_3}\Gamma\P{x_2,z_{3};z_{4},x_4}\Gamma\P{z_{5},z_{6};z_{7},z_{8}}G\P{z_{7},z_{1}}\dot{G}\P{z_{2},z_{5}}G\P{z_{8},z_{3}}G\P{z_{4},z_{6}}\\
&-4\sum_{z_1\cdots z_{8}}\Gamma\P{x_1,z_{1};z_{2},x_3}\Gamma\P{x_2,z_{3};z_{4},x_4}\Gamma\P{z_{5},z_{6};z_{7},z_{8}}\dot{G}\P{z_{7},z_{1}}G\P{z_{2},z_{5}}G\P{z_{8},z_{3}}G\P{z_{4},z_{6}}\\
&+2\sum_{z_1\cdots z_{8}}\Gamma\P{x_2,x_1;z_{1},z_{2}}\Gamma\P{z_{3},z_{4};x_3,z_{5}}\Gamma\P{z_{6},z_{7};z_{8},x_4}G\P{z_{1},z_{4}}G\P{z_{2},z_{6}}G\P{z_{8},z_{3}}\dot{G}\P{z_{5},z_{7}}
 \big)\;.
\end{split}
\end{equation}}
\end{widetext}
The first term remains unaffected by antisymmetrization operator and is the same as in eq.~\eqref{eq:Gamma4Katanin}.
The result can in principle be extended to any order, tough the computational effort grows rapidly.

\section{A stationary point formulation of the Schwinger-Dyson equations}
\label{sec:functional}
\noindent
We return to Eq.~\eqref{eq:theta4}, set $\Gamma^{(4)}-v =\Theta$ and study $\Theta:=\Theta(G,\Gamma^{(4)})$ as a functional depending on $\Gamma^{(4)}$ and $G$. 
For the solution of the SDE, $G$ itself depends on $C$ and $\Sigma$, so that the equations for $G$
and $\Gamma^{(4)}$ are really coupled, but we now consider $\Gamma^{(4)}$ and $G$ as two independent variables. To avoid confusion, the solutions of the SDE will be hatted from now on, 
i.e.\ denoted as $\hat{G}$ and $ \hat{\Gamma}^{(4)}$.
The functional $\Theta$ can be written as a gradient with respect to $\Gamma^{(4)}$. The integrability of $\Theta$ is a nontrivial property and rather interesting. It allows us to formulate the Schwinger-Dyson equations in term of a stationary point problem as will be shown below.

\begin{figure}[!htbp]
\includegraphics{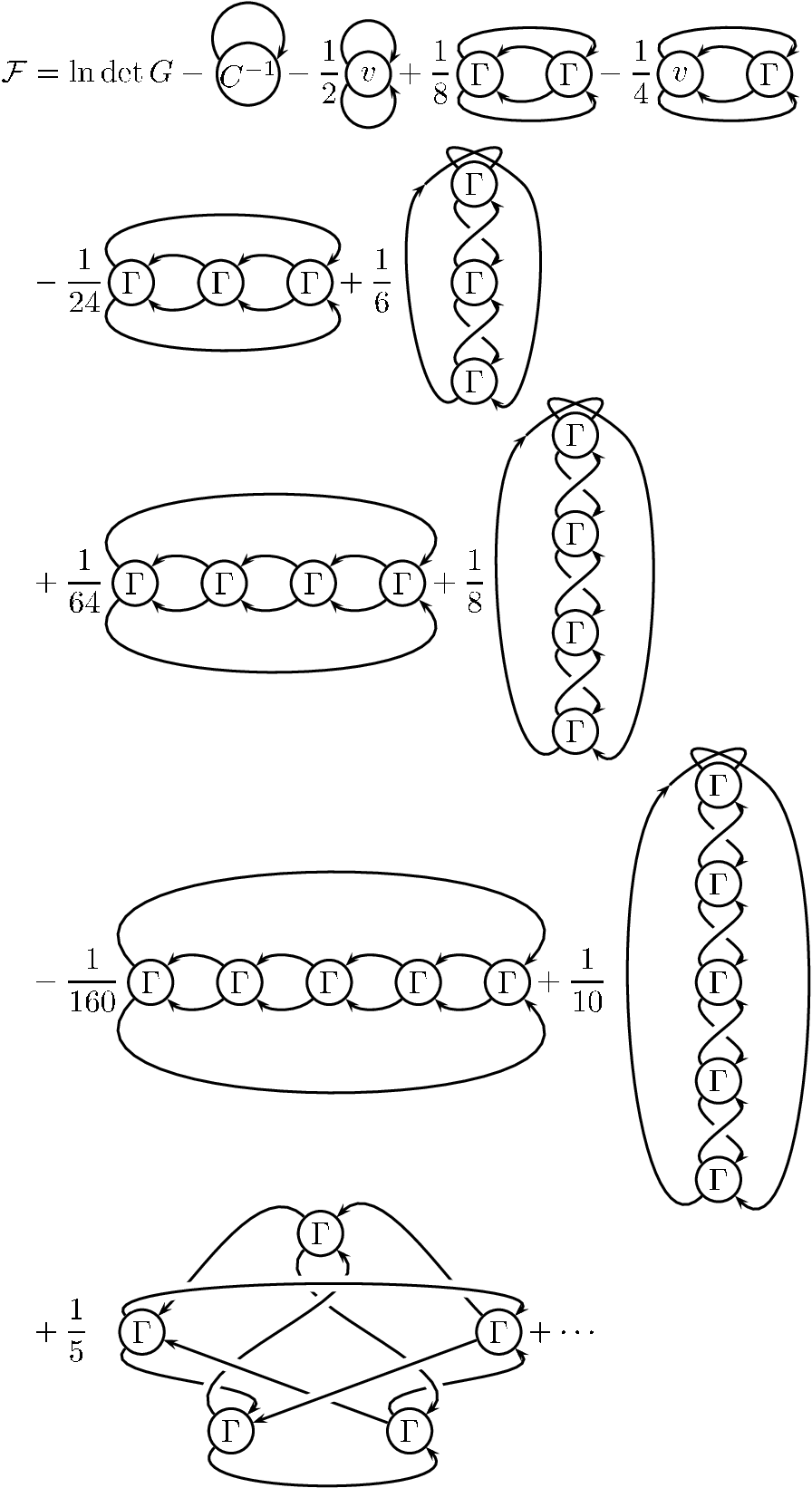}
\caption{ The diagrammatic representations of the functional $\mathcal{F}$ defined in
eq.~\eqref{eq:F}.}.
\label{fig:F}
\end{figure}

For a four-point functions $f$ define $\mathcal{C}$ as the operations
\begin{equation}
\label{eq:Ccirc}
\mathcal{C}(f) = \sum_{x,y} 
f\P{x,y;y,x} 
\end{equation}
which consists of closing the diagram and results in a scalar. Then the SD eq.~\eqref{eq:theta4} is equivalent to $\frac{\Ktd}{\Ktd \Gamma^{(4)}} \mathcal{F}_1( G,\Gamma^{(4)}) = 0$ with
\vbox{
\begin{equation}
\label{eq:F1}
\begin{split}
&\mathcal{F}_1\Kbr{G,\Gamma^{(4)}} = \frac{-1}{4} \mathcal{C} \Bigg\{ 
\frac12 \Kbr{ \Gamma^{(4)} \Kcirc \Gamma^{(4)}}\\
&-\Kbr{ \Gamma^{(4)} \Kcirc v}\\
&- \frac16 \Kbr{ \Gamma^{(4)} \Kcirc \Gamma^{(4)} \Kcirc \Gamma^{(4)}}\\
&-\frac23 \Kbr{ \Gamma^{(4)} \Kwedge \Gamma^{(4)}\Kwedge \Gamma^{(4)}} \\
& +\frac{1}{16} \Kbr{ \Gamma^{(4)} \Kcirc \Gamma^{(4)}\Kcirc \Gamma^{(4)} \Kcirc \Gamma^{(4)} }\\
&-\frac{1}{2} \Kbr{ \Gamma^{(4)} \Kwedge \Gamma^{(4)}\Kwedge \Gamma^{(4)}\Kwedge \Gamma^{(4)}}\\
& -\frac{1}{40} \Kbr{ \Gamma^{(4)} \Kcirc \Gamma^{(4)}\Kcirc \Gamma^{(4)} \Kcirc \Gamma^{(4)}\Kcirc \Gamma^{(4)}} \\
& -\frac{2}{5} \Kbr{\Gamma^{(4)} \Kwedge \Gamma^{(4)}\Kwedge \Gamma^{(4)} \Kwedge \Gamma^{(4)} \Kwedge \Gamma^{(4)}}\\
& +\frac45 \Kbr{ Q(G, \Gamma^{(4)}) \Kcirc \Gamma^{(4)} }
\Bigg\}+\mathcal{O} \Kbr{\Gamma^{(4)}}^6\;.
\end{split}
\end{equation}
}
Note that the components of the gradient with respect to $\Gamma^{(4)}$
are already antisymmetric. More precisely, we restrict $\Gamma^{(4)}$ to have the desired antisymmetry, meaning that
the components of $\Gamma^{(4)}$ are not independent. The total derivative of
a functional $F(\Gamma^{(4)})$ with 
respect to $\Gamma^{(4)}\P{x_1,x_2;x_3,x_4}$ is then given by
\vbox{
\begin{equation}
\begin{split}
\frac{\Ktd F(\Gamma^{(4)})}{\Ktd \Gamma^{(4)}\P{x_1,x_2;x_3,x_4}} & = 
\frac{\Kfd F(\Gamma^{(4)})}{\Kfd \Gamma^{(4)}\P{x_1,x_2;x_3,x_4}}
-\frac{\Kfd F(\Gamma^{(4)})}{\Kfd \Gamma^{(4)}\P{x_2,x_1;x_3,x_4}}\\
&-\frac{\Kfd F(\Gamma^{(4)})}{\Kfd \Gamma^{(4)}\P{x_1,x_2;x_4,x_3}}
+\frac{\Kfd F(\Gamma^{(4)})}{\Kfd \Gamma^{(4)}\P{x_2,x_1;x_4,x_3}}\;.
\end{split}
\end{equation}}
The factor $1/4$ in eq.~\eqref{eq:F1} is the same as the $1/4$ hidden in the definition of the antisymmetrization operator in eq.~\eqref{eq:theta4}. The  stationary point of $\mathcal{F}_1(\hat G,\Gamma^{(4)})$ is already a solution of the Schwinger-Dyson eq.~\eqref{eq:theta4}. In the next step we want to
use $\mathcal{F}_1$ to define a new functional $\mathcal{F}(\Sigma,\Gamma^{(4)})$ whose stationary point is a solution of both eqs.~\eqref{eq:theta4} and \eqref{eq:Self-Energy}.
Considering $G$ as a function of the self energy $\Sigma$ (since $G=(C^{-1}-\Sigma)^{-1}$), and let $\tilde \Gamma^{(4)}(G)$ denote a solution of the equation $\frac{\Ktd}{\Ktd \Gamma^{(4)}} \mathcal{F}_1(G,\Gamma^{(4)}) = 0$ for a given $G$. We take the derivative of
$\mathcal{F}_1\Kbr{G(\Sigma),\tilde \Gamma^{(4)}(G)}$ with respect to $\Sigma$ and make the following 
helpful observation,
\vbox{
\begin{equation}
\label{eq:helpful}
\begin{split}
&\Kbr{\frac{\Ktd}{\Ktd \Sigma\P{x_1',x_1}}\mathcal{F}_1}\Kbr{G(\Sigma),\tilde \Gamma^{(4)}(G(\Sigma))} = \\
&-\frac{1}{2}
\sum_{\begin{matrix}
\scriptstyle z_1\cdots z_4\\ 
\scriptstyle y_1\cdots y_4
\end{matrix}} \Big(
 G\P{x_1,z_1}
v\P{z_1,z_2,z_3,z_4}  G\P{z_3,y_1}  G\P{z_4,y_2}
\tilde \Gamma^{(4)}(G(\Sigma))\P{y_1,y_2,y_3,y_4}\cdot\\
&\qquad\qquad\qquad  G\P{y_4,z_2} G\P{y_3,x_1'}   \Big) +\mathcal{O} \Kbr{\Gamma^{(4)}}^6\;.
\end{split}
\end{equation}
}
The right hand side looks very similar to the last term of Eq.\ \eqref{eq:DSG}. If we define $\mathcal{F}_2$ as \\
\vbox{
\begin{equation}
\begin{split}
\mathcal{F}_2 & (G(\Sigma)) = - \sum_{z_1,z_2} G\P{z_2,z_1} \Kbr{C^{-1}}\P{z_1,z_2}\\
&-\frac12 \sum_{z_1,\dots,z_4} v\P{z_1,z_2;z_3,z_4} G\P{z_3,z_1} G\P{z_4,z_2}\\
&+\ln\Kbr{\det G}\;,
\end{split}
\end{equation}
}
and add this $\Gamma^{(4)}$-independent term to $\mathcal{F}_1$, the stationary point of
\begin{equation}
\label{eq:F}
\mathcal{F}\Kbr{\Sigma,\Gamma^{(4)}}:=\mathcal{F}_1\Kbr{G(\Sigma),\Gamma^{(4)}}+\mathcal{F}_2 (G(\Sigma))
\end{equation}
with respect to $\Sigma$ for some given $\Gamma^{(4)}$ is a solution of eq.~\eqref{eq:Self-Energy}.
Since $\mathcal{F}_2$ is independent of $\Gamma^{(4)}$ we conclude that the solution of Schwinger-Dyson equations \eqref{eq:Self-Energy} and \eqref{eq:theta4} is a stationary point of $\mathcal{F}$
\vbox{
\begin{equation}
\begin{split}
\frac{\Ktd}{\Ktd \Sigma}\mathcal{F}\Kbr{\Sigma,\Gamma^{(4)}}&=0 \\
\frac{\Ktd}{\Ktd \Gamma^{(4)}}\mathcal{F}\Kbr{\Sigma,\Gamma^{(4)}}&=0 \;.
\end{split}
\end{equation}}
The schematic representation of $\mathcal{F}$ is shown in Figure \ref{fig:F}.

\begin{acknowledgments}
This work was supported by DFG via the research group FOR 723.
\end{acknowledgments}

\FloatBarrier
\bibliography{sderg}

\end{document}